# Practical Insights into Designing Context-Aware Robot Voice Parameters in the Wild

Amy Koike
University of Wisconsin-Madison
Madison, USA
ekoike@wisc.edu

Yuki Okafuji
CyberAgent
Tokyo, Japan
okafuji_yuki_xd@cyberagent.co.jp

Sichao Song
CyberAgent
Tokyo, Japan
song_sichao@cyberagent.co.jp

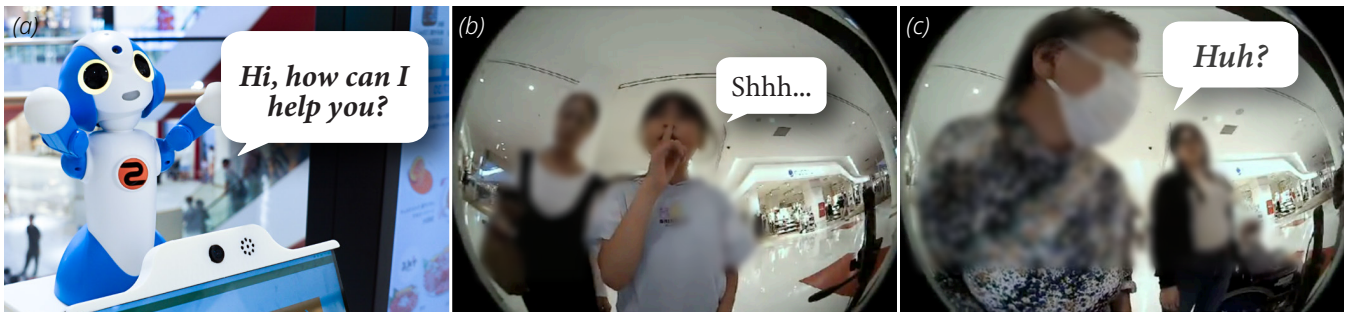

Figure 1: In this study, (a) we deployed an autonomous conversational robot in a shopping mall and investigated how contextual factors influence user experience with the robot in the wild, especially sensitivity to the robot's voice parameters (*i.e.,* speech volume and speech rate). The scenes from field deployment illustrate one of our key findings: (b) a user who interacted with the robot alone, feeling the robot sounds "too loud" and showing "shh" gesture; (c) users who interacted with the robot with the group, feeling the robot sounds "too quiet" to hear its voice and leaning their ear forward with a puzzled look.

## Abstract

Voice is an essential modality for human-robot interaction (HRI). The way a robot sounds plays a central role in shaping how humans perceive and engage with it, influencing factors such as intelligibility, understandability, and likability. Although prior work has examined voice design, most studies occur in controlled labs, leaving uncertainty about how results translate to real-world settings. To address this gap, we conducted two naturalistic deployment studies with a guidance robot in a shopping mall: (1) in-depth interviews with six participants, and (2) an eight-day field deployment using a 3×3 design varying speech rate and volume, yielding 725 survey responses. Our results show how real-world context shapes voice perception and inform adaptive, context-aware voice design for social robots in public spaces.

## 1 Introduction

Voice is one of the most important modalities for verbal human-robot interaction (HRI). The way a robot "sounds" plays a critical role in shaping how humans perceive and interact with it. Prior research has demonstrated that voice parameters, such as pitch, speaking rate, and volume, can significantly influence user perceptions of robots, such as their trust [*e.g.,* 2, 41, 42], intelligibility [11, 25], and overall likability [*e.g.,* 37].

While much of this work has been conducted in controlled environments, human-robot interactions that occur in the real world often involve unexpected, complex, and noisy situations. These "in-the-wild" contexts, such as shopping malls, airports, or hospitals, introduce challenges that are rarely captured in laboratory-based studies. For instance, unpredictable environmental noise, diverse user populations, and spontaneous multi-party interactions can all shape how people experience a robot's voice. However, there remains a limited understanding of how these real-world factors influence user perceptions of a robot's voice design.

To address this gap, we explore how voice parameters affect human perception of robots in public spaces. Specifically, we ask this question: how do contextual factors modulate sensitivity to a robot's speech rate and volume in the wild?

Our research makes the following key contributions: **(1) In-depth interviews:** we conducted a series of interviews in a shopping mall where participants interacted with a guidance robot, identifying contextual factors and voice parameters that shape user perceptions of robot voice design; **(2) Field study:** we systematically evaluated robot voice design in the wild by testing a 3 × 3 factorial combination of speech rate and volume; **(3) Design implications:** we present empirical evidence from both studies and derive design implications for creating more effective, user-centered robot voice designs in real-world environments.

## 2 Related Work
### 2.1 Voice Design and Perception in HRI

Extensive research in human-robot interaction (HRI) has shown that a robot's vocal qualities strongly influence social judgments, including trust [*e.g.*, 2, 41, 42], perceived competence [*e.g.*, 11, 16], user expectations [*e.g.*, 10, 25], and likability or willingness to engage [*e.g.*, 15, 37]. These findings highlight the importance of carefully and deliberately designing voice for social robots.

Past work has identified several key parameters that influence users' perceptions and interactions with robots. Pitch has been shown to affect perceptions of warmth, authority, and trustworthiness, with higher pitch often associated with friendliness and lower pitch with competence and dominance [26, 27, 38]. Speech rate plays a role in balancing intelligibility and perceived intelligence: slower speech can improve comprehension but may reduce perceived expertise, whereas faster speech can signal competence while increasing cognitive load [11, 25]. Volume is especially relevant for robots deployed in public or noisy spaces, where inappropriate loudness can hinder engagement or create discomfort [24, 33].

Beyond the design of the robot's voice, prior research has shown that characteristics of the listener also play a significant role in shaping perceptions of the robot. For instance, differences in gender, age, and even cultural background have been found to influence how users evaluate and respond to a robot's voice [6, 7, 9, 12, 15, 38]. These findings motivated us to consider further how contextual factors, especially for in-the-wild deployment, affect perceptions of robot voices, which we elaborate on in the following subsection.

### 2.2 From Lab to the Wild

While "HRI in the wild" has been an active research area within the HRI community [*e.g.*, 4, 18, 34–36, 44], many HRI observation are still conducted in controlled laboratory settings.

Although controlled settings ensure rigor, translating laboratory findings to real-world HRI remains challenging [17]. Prior work shows that physical and social contexts strongly shape how people interact with robots [5, 23, 29], so lab behaviors may not reflect field behavior. Building on Koike et al. [20], which identified four user motivations in real-world robot interactions, we treat context and motivation as key cues for adapting robot voice to support more effective and pleasant HRI in the wild.

### 2.3 Adaptive, Context-Aware Robot Voices

Recent research has explored ways to make robot voices adapt to their surroundings and users, instead of relying on fixed voice settings. A key focus has been adjusting a robot's speech to remain clear and appropriate in real-world, noisy environments. For instance, Bui and Chong [3] developed an autonomous volume control system that uses deep reinforcement learning to automatically raise or lower the robot's speaking volume depending on background noise levels. Similarly, Ren et al. [33] proposed methods for improving speech intelligibility by adapting voice characteristics so the robot can be better understood even in challenging acoustic conditions. Tuttosi et al. [43] took a broader approach, outlining how robots can speak in a contextually appropriate way by adjusting various speech features based on environmental and social cues. Other studies have focused on adapting specific vocal features, such as pitch and prosody. Lubold et al. [22] showed that changing pitch patterns can make a robot's speech sound more responsive and socially appropriate. Researchers have also explored personalized voice adaptation for particular groups of users. For example, Panduwawala et al. [31] designed adaptive voice strategies for children with autism spectrum disorder (ASD), helping the robot match the child's needs and comfort level. Tielman et al. [40] studied how robots can adjust emotional expressions in their voices to better support children in interactive tasks.

Overall, these studies demonstrate that adapting a robot's voice, whether by adjusting volume, pitch, speech clarity, or emotional tone, can enhance user experience across various settings and populations. However, most prior work has been tested in controlled laboratory studies, so little is known about how to apply these findings to in-the-wild human-robot interaction. The ultimate goal of this paper is also to develop and build adaptive, context-aware robot voices. To address this, we focus on the types of contextual cues that can impact the user's perception of robots in the wild.

## 3 Preliminary Interview

To understand how users perceive and engage with the robot in the wild, we conducted a series of in-depth, semi-structured interviews. The focus of these interviews was to investigate the design space of robot voice in the wild, including what kind of contextual cues can shape user experience, as well as identify design spaces of robot voice parameters that may create a better HRI experience. All research activities were reviewed and approved by the Ethical Review Board of Osaka University.

### 3.1 Participants

We recruited six participants (three male, three female), aged 21–44 years ($M = 31.83$, $SD = 9.33$), through a temporary staffing agency by sending a recruitment request via email. All participants were located in Japan and were fluent speakers of Japanese. The study procedure lasted approximately four hours, and participants were compensated at a rate of 1,800 JPY per hour. The interviews were conducted in two sessions, with three participants in each session.

### 3.2 Interview Setting

**Robot and System.** We used a humanoid robot named Sota for our study (Figure 2(b)). Sota robots have been widely employed in various HRI studies as service robots [*e.g.*, 1, 20, 39]. Sota features a torso, head, and two arms, allowing it to perform body gestures. Sota's main task was to assist shoppers by providing guidance and information. To complement Sota's verbal interactions, we integrated a 13-inch display that shows recognized text and content synchronized with Sota's current state. Both the robot and the display were mounted on a 3D-printed container, which also housed a 180-degree fisheye camera, microphone, speaker, and cables. The system integrates vision, speech, and dialogue management to provide interactive guidance in a shopping mall setting. A 180-degree fisheye camera and PoseNet [19] are used to detect visitors and estimate their proximity to the robot, while a microphone and Google Speech-to-Text API handle speech input. Based on the visitor's distance, the system transitions between three states: no visitor present, visitor detected, and visitor ready to interact. Speech recognition is activated only when a visitor is sufficiently close to interact, thereby reducing errors in noisy environments. When a visitor is detected, the robot greets them using template utterances and, once engaged, processes the input to generate responses. The dialogue system combines rule-based and generative actions. Rule-based actions handle predictable interactions, such as greetings or state changes, while GPT-4o [30] generates context-aware responses using the shopping mall information and dialogue history. If a user requests directions, the system triggers a navigation function to display an animated route on a screen while providing a spoken explanation. The robot, Sota, enhances interaction through synchronized multimodal feedback, including speech synthesis, LED indicators, and gesture animations. Predefined gestures are paired with specific words, and subtle random motions signal that the robot is active. Together, these features create a responsive and natural interaction experience for users navigating public spaces.

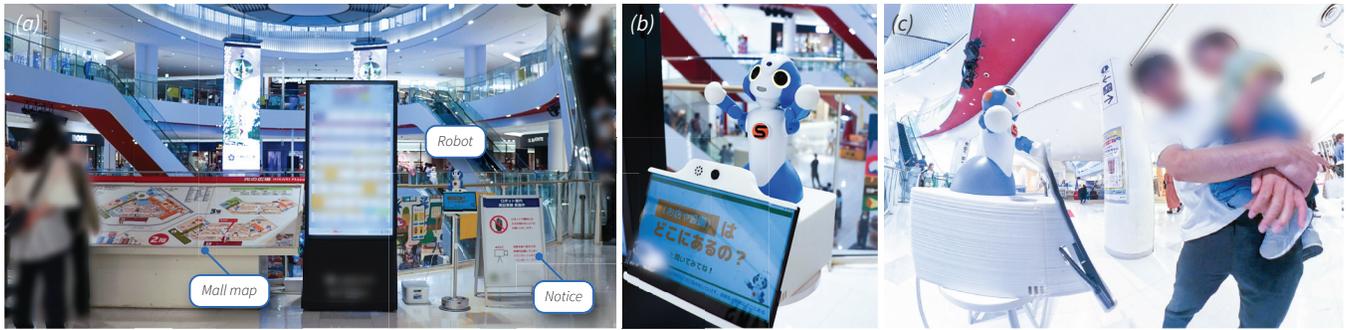

Figure 2: Study Setting. (a) The system was deployed on the 2nd and 3rd floors of a shopping mall, positioned next to a mall map. A video camera was installed behind the robot, and an experimental notice was placed beside the system; (b) The Sota robot, a 27 cm-tall tabletop social humanoid, is performing bodily gestures; (c) scene from the field study.

**Procedure.** The interviews were conducted at a shopping mall in Osaka, Japan. A guidance robot was placed near the mall directory map to provide guidance and information about the shopping mall. Three researchers facilitated the study: two were present on-site, while one participated remotely. Additionally, two research assistants assisted with participant coordination and logistics for the interviews. Each interview session lasted approximately four hours and consisted of two main stages: an experience component and an interview component. The experience stage comprises the following activities: natural interaction, role-playing scenarios, and group interaction. In the natural interaction, participants were first encouraged to approach and interact with the robot freely, simulating a natural encounter. Each participant engaged with the robot individually while the others observed. At the role-playing, participants were asked to engage in structured role-playing tasks using pre-defined scenarios. This allowed the researchers to observe how participants interacted with the robot in various social contexts. Finally, in the group interaction, all participants were asked to approach and use the robot together to explore how group dynamics influenced perceptions of the robot's voice and behavior. During these activities, the remote researcher observed participants through the robot's camera feed, while the two on-site researchers monitored interactions from a distance to avoid influencing behavior. After the interaction phase, participants and researchers relocated to a private interview setting. Participants were asked to reflect on their experiences interacting with the robot, including their overall impressions, any discomfort or inconvenience they encountered, their preferences and perceptions of the robot's voice, and relevant contextual factors influencing those reflections.

### 3.3 Findings

The interviews were recorded and transcribed for analysis. Given the exploratory nature of the study, we adopted a flexible, informal approach rather than a structured qualitative methodology. The research team highlighted key quotes and engaged in iterative discussions to identify emerging themes, with a focus on design parameters and consideration related to robot voices in the wild. Below, we summarize the key findings.

**Time of Day and Environmental Noise.** Participants noted that their preferences varied depending on the time of day and environmental conditions. For example, weekends were described as generally noisier, such as in shopping malls or other public spaces, making it harder to hear the robot. In these situations, a louder voice was preferred to ensure intelligibility, whereas quieter weekday environments required less volume.

**Presence of Bystanders.** The presence of bystanders strongly influenced voice preferences. When others were waiting behind the participant to use the robot, or when a group of people was nearby, participants preferred the robot to speak more quietly and deliver short, direct instructions to minimize social discomfort. This preference stemmed from a desire to finish the interaction as quickly as possible to avoid getting attention or holding up others waiting to use the robot. In addition, this preference was pronounced when the information being exchanged was potentially private or sensitive; for instance, when asking for directions to an underwear store.

**Social Context: Alone vs. Accompanied.** The social dynamics of the interaction also played a role. Participants reported feeling more self-conscious and embarrassed when interacting with a robot alone in an open space, especially in highly visible locations. When accompanied by friends or family, they felt more at ease and less concerned about drawing attention to themselves.

**Verbal Engagement.** In socially noticeable settings, users may feel embarrassed and prefer brief interactions, making them less sensitive to vocal parameters (prior work also showed similar findings [28]). Once engagement deepens, however, users become more sensitive to fine-grained vocal features—such as volume, rate, and timing—and expect more adaptive behaviors. Thus, both whether and how deeply users engage should guide how voice parameters are set across interaction phases.

**Voice Audibility and Attention.** While a quieter voice can reduce social awkwardness, participants noted a trade-off: if it was too quiet, users struggled to maintain attention and sometimes abandoned the interaction. This tension highlights the need to balance privacy and audibility. Participants also discussed directional speakers, which could improve privacy but might make users appear to be talking to themselves, increasing embarrassment. They suggested using clear visual or explicit cues to signal when the robot is speaking to mitigate this issue.

These findings suggest that users' preferences for robot voices are situational, shaped by social, spatial, and environmental factors. From the findings, we defined the following elements as new important context factors and analyzed them for our subsequent field deployment study, as described in Section 4: Time of Day, Presence of Bystanders, Group Level, and Presence of Verbal Engagement.

## 4 Field Experiment

We conducted a field study to investigate how contextual factors interact with two voice parameters–volume and speech rate–during human-robot verbal interaction in a real-world setting (Figure 2). All research activities were reviewed and approved by the Ethical Review Board of Osaka University.

### 4.1 Study Setting

**Robot and System.** We deployed the same robot and system used in the preliminary interview. The robot was a fully autonomous conversational system equipped with speech recognition and natural language processing capabilities, designed to assist shoppers by providing guidance and information about a shopping mall.

**Procedure.** Our field study was conducted at a shopping mall located in Osaka, Japan, the same location as the preliminary interview, over eight consecutive days in May and June 2025. The robot was positioned next to a mall map on both the second and third floors to maximize visibility and encourage interaction (Figure 2(a)). Data collection was balanced across weekdays and weekends and across different time periods of the day to capture a diverse range of visitor interactions. The system operated for eight hours per day. When a visitor initiated an interaction with the robot, the system autonomously engaged in conversation. After the interaction ended, the experimenter approached the participant, explained the purpose of the study, and invited them to complete a brief questionnaire measuring their perceptions of the robot's voice characteristics.

### 4.2 Stimulus Design

The stimuli were constructed using a 3 × 3 factorial design, varying two parameters: voice volume (V1–3) and speech rate (S1–3). We selected volume and speech rate as voice parameters based on the preliminary interviews, which highlighted their impact on user experience with a robot in the wild. For example, participants explained that if the volume were too loud, they might feel embarrassed to speak to it, and if they were in a hurry, they preferred the robot to speak faster. Pitch as a voice parameter is also known as one of the design factors, however we did not include it because changes in pitch were mentioned less frequently in the preliminary interviews. In addition, because our study used a between-subjects design, subtle pitch differences would not be directly compared within the same listener, making perceptual differences more difficult to detect. For these reasons, pitch was excluded from the stimulus parameters. For voice volume, three levels were defined (V1, V2, V3), based on prior literature and the maximum allowable volume (V3) discussed with the shopping mall manager. The environmental noise level in the area, excluding the robot, typically ranged from 65 to 70 dB on both weekdays and weekends when the robot was deployed. Based on this, the levels were determined as follows: V1 as S/N ratio of +3, resulting in approximately 70–73 dB; V2 as S/N ratio of +9 to +10, resulting in approximately 79–80 dB; and V3 as S/N ratio of +13, resulting in approximately 83–85 dB. For speech rate, three levels were defined (S1, S2, S3) based on relevant guidelinesand the researcher's prior experience. Among these, S2 reflected the researcher's usual or baseline setting, while S1 (slower) and S3 (faster) were determined through discussions within the research team to ensure a clearly distinguishable difference between levels. The levels were implemented using the Google Text-to-Speech (TTS) system, where the speech rate parameter acts as a multiplier of the default speed. Each level was defined as follows: S1 at 0.8 (approximately 120 WPM), S2 at 1.2 (approximately 150 WPM), and S3 at 1.6 (approximately 230 WPM).

### 4.3 Data Collection and Measures

All interactions between the robot and users were video recorded. Users were informed that video recordings would be collected and that their consent to the use of their videos in the research was obtained via instructions posted next to the robot. At least one experimenter was present in the space to address any issues. Although the experimenter stood slightly apart from the system, they continuously monitored the system and interactions. During the study, when an interaction occurred, the experimenter approached the participants and asked them to complete a short questionnaire. In total, we obtained 773 responses from the participants, however we removed 48 responses due to issues in the recordings, such as footage that was not properly captured or did not clearly show the interaction with the robot. Because we wanted to code user behaviors in video recordings, these responses were excluded from the analysis. As a result, 725 valid responses were analyzed.

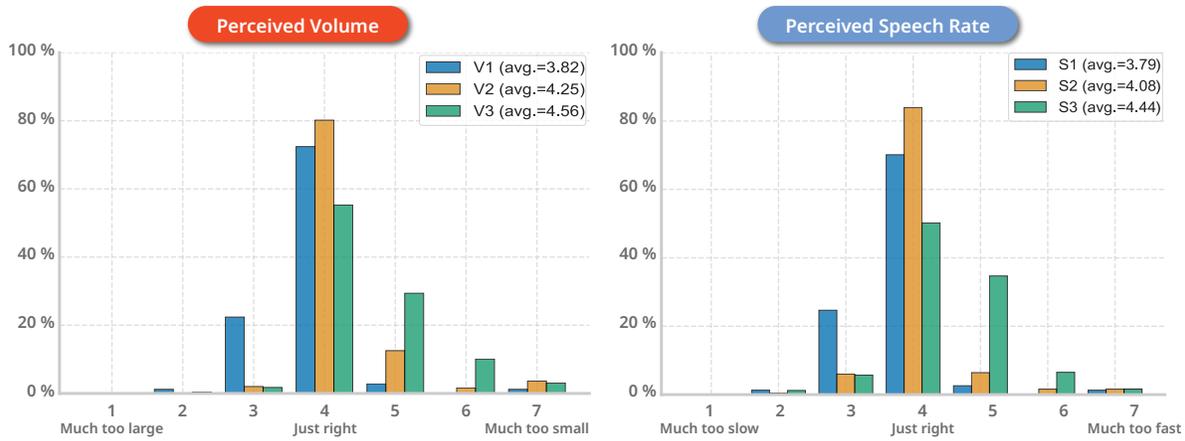

Figure 3: Distributions of subjective ratings by manipulation level. Left: perceived volume $Q_V$ across V1–V3. Right: perceived speech rate $Q_S$ across S1–S3.

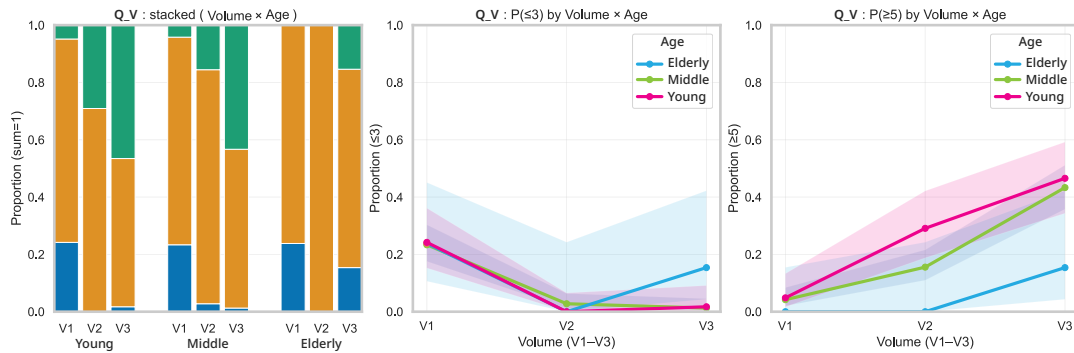

Figure 4: Perceived volume $Q_V$ by Volume (V1–V3) × Age (Young / Middle / Elderly). The left panel shows $Q_V$ response distributions as a function of volume condition and age group. Bars are normalized to sum to 1, and stacked colors represent the proportion of responses in each perceived-volume category. The right two panels plot model-predicted probabilities for extreme judgments: P($Q_V < 3$) ("too small") and P($Q_V > 5$) ("too large").

**Dependent variables.** The primary outcome measures were the participants' *perceived volume* $Q_V$ and *perceived speech rate* $Q_S$ of the robot's voice. These were assessed using a seven-point Likert scale (1 = much too small/much too slow, 7 = much too large/much too fast). This subjective evaluation captured the participants' immediate impressions of the robot's voice.

**Independent variables.** Several demographic and contextual factors extracted from previous studies and the preliminary interview were coded to the video recordings by trained researchers and annotators: *Age:* categorized perceived age of the participants from young, middle-aged, or elderly (due to age-related differences in auditory abilities [14]); *Day:* whether the interaction occurred on a Weekday or Weekend; *Bystanders:* whether bystanders were present during the interaction (True/False); *Group:* whether the participant interacted alone or as part of a group; *Verbal Engagement:* whether the participant engaged in extended conversation with the robot beyond basic interaction (True/False); and *Motivation:* categorized as Function (goal-driven, minimal dialogue to complete a task), Experiment (interaction to try the robot's capabilities), Curiosity (interest-driven, playful or observational), or Education (caregiver-mediated interactions aimed at children's experience) [20]. Two researchers (R1 and R2) led this coding process, and four coders analyzed the video recordings. Age, Group, Presence of Observers, Level of Verbal Engagement, and Day were annotated by the four coders after thorough training by R2. Discrepancies were resolved through discussion among the coders and the researchers to ensure reliability. For user motivation, R1 coded 57% of the data while R2 coded the remainder. Cohen's Kappa was calculated to measure the agreement between the two raters by using 10% of the data. The reliability was sufficiently high ($\kappa$ = 0.78).

### 4.4 Analysis

Because the two outcome variables ($Q_V$: perceived volume, $Q_S$: perceived speech rate) were measured on a seven-point Likert scale, they were treated as ordinal rather than continuous. Accordingly, we used cumulative link models (CLMs) with a proportional-odds logistic link [8]. CLMs estimate how predictors and their interactions influence the probability that a response falls at or above a given category, making them well-suited for questions such as whether certain conditions reduce perceptions of being "too loud" or "too fast" while preserving judgments of being "just right."

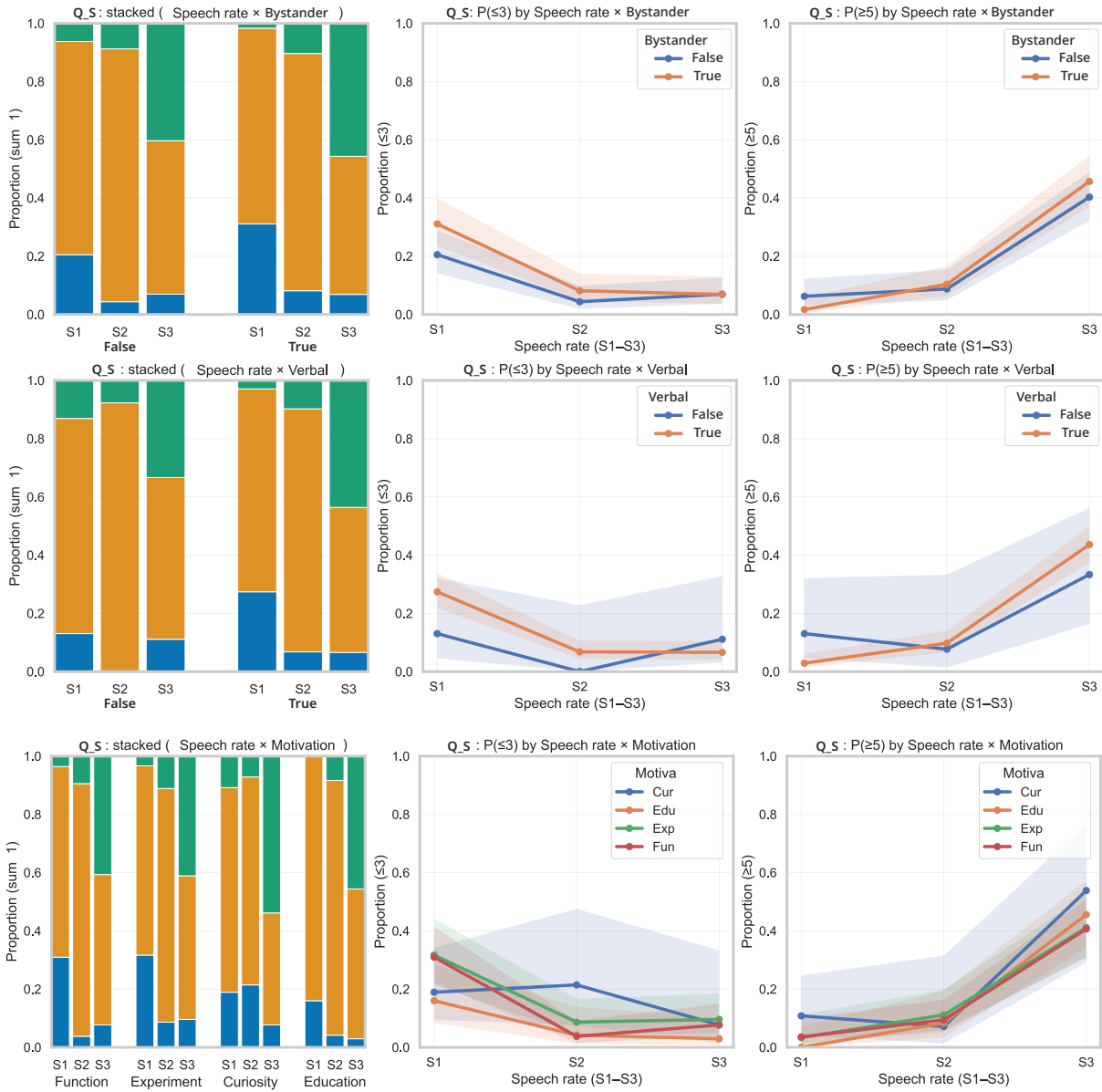

Figure 5: Perceived speech rate $Q_S$ by Speech Rate (S1–S3) × Context (Bystanders / Verbal engagement / Motivation). The left panels show $Q_S$ response distributions. Bars are normalized to sum to 1, and stacked colors represent the proportion of responses in each perceived speech rate category. The right panels plot P($Q_S$ < 3) ("too slow") and P($Q_S$ > 5) ("too fast").

All categorical predictors were coded using effect coding (sum-to-zero), so the intercept represents the grand mean on the logit scale, and each coefficient reflects deviation from that mean. Ordered predictors (*Volume V*, *Speech rate S*, and *Age*) were modeled using orthogonal polynomial contrasts: the linear term (.L) captures monotonic trends (*e.g., V1→V3*), and the quadratic term (.Q) captures curvature (U- or ∩-shaped effects).

Separate models were estimated for perceived volume ($Q_V$) and perceived speech rate ($Q_S$). In the $Q_V$ model, we included interactions between voice volume and each contextual factor; in the $Q_S$ model, we examined interactions between speech rate and the same contextual factors. The full model formulas were:

Table 1: Sample characteristics and $V \times S$ cell sizes ($n$=725). Percentages are relative to the total sample. F, Ex, C, and Ed refers to *Function*, *Experiment*, *Curiosity*, and *Education*.

| Characteristic | n (%) |
| --- | --- |
| V1-S1 / V1-S2 / V1-S3 | 85 (11.7) / 80 (11.0) / 85 (11.7) |
| V2-S1 / V2-S2 / V2-S3 | 85 (11.7) / 83 (11.4) / 79 (10.9) |
| V3-S1 / V3-S2 / V3-S3 | 61 (8.4) / 86 (11.9) / 81 (11.2) |
| Young / Middle / Old | 175 (24.1) / 504 (69.5) / 46 (6.3) |
| Weekday / Weekend | 542 (74.8) / 183 (25.2) |
| Bystanders (True / False) | 370 (51.0) / 355 (49.0) |
| Solo / Group | 84 (11.6) / 641 (88.4) |
| Verbal (True / False) | 671 (92.6) / 54 (7.4) |
| Motivation (F / Ex / C / Ed) | 281 (38.8) / 214 (29.5) / 64 (8.8) / 166 (22.9) |

$$Q_V \sim V + S + Age + Day + Gro + Bys + Ver + Mot + \quad (1)$$
$$V{:}S + V{:}Age + V{:}Day + V{:}Gro + V{:}Bys + V{:}Ver + V{:}Mot$$
$$Q_S \sim V + S + Age + Day + Gro + Bys + Ver + Mot + \quad (2)$$
$$S{:}V + S{:}Age + S{:}Day + S{:}Gro + S{:}Bys + S{:}Ver + S{:}Mot$$

In these formulas, the symbol ":" denotes an interaction term. For example, "S:V" represents the interaction between S and V, indicating that the effect of S on the outcome may vary depending on the level of V. The variables are defined as follows: Age (participant age), Day (weekday or weekend), Gro (solo or group participation), Bys (presence of bystanders), Ver (presence of verbal interaction), and Mot (participant motivation pattern).

**Assumption checking.** We evaluated the proportional-odds (PO) assumption using nominal tests. When a violation was detected, we fit partial proportional-odds models (PPOMs) that relaxed PO for the offending factor and compared fit via likelihood-ratio tests (LRT) and AIC.

## 5 Results

### 5.1 Sample and Allocation

As a result, we collected $n$=725 valid questionnaires. Condition cell counts for the 3 (V1-3) × 3 (S1-3) design ranged from 61 to 86 per cell. Detailed sample characteristics and cell sizes are summarized in Table 1. Descriptive histograms of $Q_V$ and $Q_S$ distributions by manipulation levels are provided in Figure 3.

### 5.2 Perceived Volume

**Model fit and assumption.** The $Q_V$ CLM showed no meaningful PO violations (nominal test non-significant). Model fit: logLik= −610.36, AIC=1300.7. See Table 2 and Figure 4 for selected statistics and the Appendix for full coefficients.

**Main effects.** The linear contrast for *Volume* was positive and significant ($p < 0.01$), confirming that higher output levels increased the odds of selecting a higher ("louder") category. The linear contrast for *Age* was negative ($p<0.01$), suggesting that older visitors were less likely to rate the robot as loud, all else equal.

**Interactions.** A significant $V \times Age$ interaction on the linear terms ($p$=0.01) indicated that increases in *Volume* produced more minor shifts toward "louder" categories for older visitors (attenuated slope from $V1{\to}V3$). Other pre-registered interactions were not significant in the CLM.

Table 2: $Q_V$ (perceived volume): Selected significant CLM effects (odds ratios, OR).

| Effect | β | SE | p | OR [95% CI] |
| --- | --- | --- | --- | --- |
| V.L (linear) | 1.47 | 0.45 | < 0.01 | 4.33 [1.78, 10.52] |
| Age.L (linear) | -1.15 | 0.31 | < 0.01 | 0.32 [0.17, 0.58] |
| V.L × Age.L | -1.26 | 0.49 | 0.01 | 0.28 [0.11, 0.74] |

Table 3: $Q_S$ (perceived speech rate): Selected significant CLM effects (odds ratios, OR). Motivation (F / Ex / C / Ed) refers to *Function*, *Experiment*, *Curiosity*, and *Education*, respectively.

| Effect | β | SE | p | OR [95% CI] |
| --- | --- | --- | --- | --- |
| S.L (linear) | 0.98 | 0.42 | 0.02 | 2.67 [1.18, 6.04] |
| S.L × Bys (True vs. False) | 0.31 | 0.15 | 0.03 | 1.36 [1.02, 1.82] |
| S.L × Ver (True vs. False) | 0.70 | 0.31 | 0.02 | 2.02 [1.11, 3.68] |
| S.Q × Mot (F vs. Ex) | -0.68 | 0.28 | 0.01 | 0.50 [0.29, 0.87] |
| S.Q × Mot (F vs. Ed) | 1.16 | 0.52 | 0.03 | 3.20 [1.15, 8.93] |

### 5.3 Perceived Speech Rate

**Model fit and assumption.** The $Q_S$ CLM yielded logLik=−636.96, AIC=1353.9. A nominal PO test suggested a violation for *Age* (LR $\chi^2$=20.13, df= 8, $p < 0.01$). A PPOM relaxing PO for *Age* improved AIC by $\Delta$AIC= −4.1, but produced a numerically singular Hessian; we therefore report CLM estimates as primary and interpret age-related findings conservatively. (See Table 3 and Figure 5 for selected statistics and the Appendix for full coefficients.)

**Main effects.** The linear contrast for *Speech rate* was positive and significant (S.L: $p$=0.02), confirming that faster settings pushed ratings toward "faster" categories.

**Interactions.** We observed four theoretically relevant interactions: (1) $S \times$ *Bystanders (linear)* ($p$=0.03); with bystanders present, increases in $S$ produced stronger shifts toward "faster" ratings; (2) $S \times$ *Verbal engagement (linear)* ($p$=0.02); visitors who engaged in extended conversation were more sensitive to rate changes; (3) $S \times$ *Motivation (quadratic, Function vs. Experiment)* ($p$=0.01); a concave pattern where increases from $S2{\to}S3$ saturated for visitors *experimenting* with the robot; (4) $S \times$ *Motivation (quadratic, Function vs. Education)* ($p$=0.03); a convex pattern where response accelerated at $S3$ for *education*-driven visitors.

## 6 Discussion

Our field experiment examined the impact of contextual factors on users' perceptions of robot voice volume and speech rate in a shopping mall. Across eight days, we collected 725 valid questionnaires distributed evenly across the nine experimental conditions (V1–3 × S1–3). This large sample provided sufficient statistical power to examine both main effects and theoretically relevant interactions.

## 6.1 Perceived Volume

A manipulation check confirmed that the volume level manipulation worked as intended. As expected, higher volume settings significantly increased the odds of selecting louder response categories, indicating that volume level strongly shaped perceived loudness. It needs to be noted that this effect was moderated by the visitor's age. Older participants were less likely to perceive the robot as loud at any given setting, and volume increases produced more minor perceptual shifts for these individuals. This aligns with prior work demonstrating age-related differences in auditory sensitivity [*e.g.,* 14], suggesting that a one-size-fits-all volume setting may not adequately serve diverse public audiences. However, while increasing a robot's volume level to accommodate older listeners can improve audibility, it can be a dual-edged intervention: louder playback elevates ambient noise for bystanders, heightens embarrassment and social visibility for the user using the robot [28], and enlarges the radius within which private or sensitive content can be overheard. Designing a tailored robot voice, therefore, demands a multi-faceted approach that balances target users' audibility against these externalities. No other interactions reached significance, indicating that the physical volume level itself was the dominant determinant of perceived loudness.

## 6.2 Perceived Speech Rate

As a manipulation check, perceived speech rate increased monotonically across S1–S3, confirming that our speech rate manipulation worked as intended; faster speech settings reliably shifted ratings toward "faster" categories, but several contextual factors modulated this relationship. First, "social pressure and attentional reallocation" played a key role in the presence of the bystanders. When bystanders were present, users often experienced subtle time pressure to "not hold up the line," which increased their sensitivity to temporal features of the robot's speech such as tempo, pause length, and articulation density [21]. In addition, in noisy public spaces, elevated semantic processing demands further lowered the threshold for judging speech as "fast," amplifying perceived speed even when the actual rate remained unchanged [32]. Second, "depth of verbal engagement" heightened sensitivity to temporal cues. Users engaged in active dialogue–rather than passive observation–relied on rhythmic elements such as response timing and turn-taking gaps [45]. As conversational exchange developed, deviations from a baseline rhythm became more salient, making even small increases in rate more readily perceived as "fast."

Finally, "user's motivation to interact with the robot" introduced non-linear dynamics. Users motivated by *Experiment* were primarily focused on trying the system itself, which yielded a wider tolerance band for speech rate and a plateauing of perceived speed at higher settings. By contrast, for *Education*-driven visitors, where parents often balanced shopping or other goals with satisfying a child's curiosity, concise and efficient presentations were valued. For these users, slightly to highly faster speech tended to be interpreted positively as transparent and efficient assistance, causing perceptions to accelerate sharply at the quickest rate tested.

These patterns suggest that speech rate is not a simple linear control, but a context- and motivation-dependent factor shaped by social setting, engagement depth, and user intent.

## 6.3 Design Implications

These findings highlight the importance of adaptive, context-aware voice design for social robots in public settings [13, 14, 33]. For volume, systems should dynamically adjust levels based on environmental noise, as suggested by prior work, and demographic factors such as age. For speech rate, designers should use sensors or interaction logs to detect social context–such as bystander presence and whether engagement is brief or extended–and adapt delivery accordingly. User motivation can further support personalization: for example, for Education-motivated users, slower rates may improve comprehension for children, while slightly faster rates may feel more efficient to accompanying adults; in casual exploration contexts (*i.e.,* Experiment users), moderate rates can help avoid overwhelming users. Together, these results provide guidance for designing robot voices that balance intelligibility, privacy, and user comfort across contexts. Although our study was conducted in a shopping mall, we expect these insights to generalize to similar public spaces such as airports and retail environments where service robots are increasingly deployed.

## 6.4 Limitations and Future Work

Our study has several limitations:

**Contextual and Cultural Generalizability.** We acknowledge potential cultural differences that may influence our findings. Although our deployment took place in Japan, it was not feasible to assess each user's cultural background during natural, walk-up interactions. Some of our findings, such as the relationships between volume and age, are consistent with prior studies and may generalize beyond this context. In contrast, the influence of bystanders may reflect culturally specific social norms in Japan, where people often adjust their behavior based on those around them.

**Measurement and Evaluation Constraints.** While the study leveraged a naturalistic setting with a large sample, the analysis of perceived speech characteristics relied on self-report ratings and a three-level manipulation scheme. Future research should incorporate objective acoustic measures and a broader range of voice parameters to more accurately capture the real-time user experience. In addition, our models treated demographic and motivational factors as categorical predictors; richer measures of individual differences (*e.g.,* hearing profiles, personality traits) could refine personalization strategies.

**Design and Analysis Scope.** We did not include voice pitch, although it likely interacts with age-related high-frequency sensitivity loss [13]; future work should test pitch across age groups. Moreover, we did not analyze interaction length or conversational content, as duration was difficult to annotate consistently due to users' movement in and out of view. Interaction length may still be meaningful, as it reflects engagement and is closely tied to content and motivation; we partly addressed this through four motivation categories, where function-motivated users tended to interact briefly, while curiosity-, experiment-, or education-motivated users often engaged longer.

## Acknowledgement

We would like to acknowledge Sari Takahashi and Aren Kagawa for their assistance in data collection and analysis.

## A  Appendix

Table 4 provides the full CLM results.

Table 4: Full Cumulative link models (CLMs) results with odds ratios and 95% CIs. Left block: Q2 (Loudness). Right block: Q3 (Speech rate). Motivation (F / Ex / C / Ed)

| | $Q_V$ (Loudness) | | | | | | $Q_S$ (Speech rate) | | | | |
|---|---|---|---|---|---|---|---|---|---|---|---|
| **Predictor** | $\beta$ | SE | z | p | OR [95% CI] | **Predictor** | $\beta$ | SE | z | p | OR [95% CI] |
| V.L | **1.47** | **0.45** | **3.24** | **0.00*** | **4.33 [1.78, 10.52]** | S.L | **0.98** | **0.42** | **2.35** | **0.02*** | **2.67 [1.18, 6.04]** |
| V.Q | -0.57 | 0.44 | -1.29 | 0.20 | 0.56 [0.24, 1.35] | S.Q | 0.47 | 0.42 | 1.11 | 0.27 | 1.59 [0.70, 3.63] |
| S.L | -0.09 | 0.16 | -0.56 | 0.58 | 0.91 [0.66, 1.26] | V.L | 0.01 | 0.16 | 0.06 | 0.95 | 1.01 [0.74, 1.37] |
| S.Q | -0.02 | 0.17 | -0.10 | 0.92 | 0.98 [0.71, 1.37] | V.Q | 0.11 | 0.16 | 0.68 | 0.50 | 1.12 [0.81, 1.54] |
| Age.L | **-1.15** | **0.31** | **-3.76** | **0.00*** | **0.32 [0.17, 0.58]** | Age.L | 0.35 | 0.27 | 1.29 | 0.20 | 1.42 [0.83, 2.43] |
| Age.Q | -0.37 | 0.19 | -1.91 | 0.06 | 0.69 [0.47, 1.01] | Age.Q | -0.00 | 0.18 | -0.00 | 1.00 | 1.00 [0.71, 1.42] |
| Day (Weekday vs. Weekend) | 0.08 | 0.13 | 0.64 | 0.52 | 1.09 [0.84, 1.40] | Day (Weekday vs. Weekend) | -0.06 | 0.12 | -0.46 | 0.64 | 0.95 [0.74, 1.20] |
| Gro (True vs. False) | 0.19 | 0.14 | 1.41 | 0.16 | 1.21 [0.93, 1.58] | Gro (True vs. False) | 0.03 | 0.14 | 0.20 | 0.84 | 1.03 [0.78, 1.35] |
| Bys (True vs. False) | 0.03 | 0.09 | 0.35 | 0.73 | 1.03 [0.87, 1.23] | Bys (True vs. False) | -0.10 | 0.09 | -1.10 | 0.27 | 0.91 [0.76, 1.08] |
| Ver (True vs. False) | 0.19 | 0.20 | 0.96 | 0.34 | 1.21 [0.82, 1.80] | Ver (True vs. False) | -0.23 | 0.19 | -1.20 | 0.23 | 0.79 [0.54, 1.16] |
| Mot (F vs. Ex) | 0.24 | 0.15 | 1.61 | 0.11 | 1.27 [0.95, 1.70] | Mot (F vs. Ex) | -0.06 | 0.15 | -0.37 | 0.71 | 0.95 [0.70, 1.27] |
| Mot (F vs. C) | -0.01 | 0.16 | -0.10 | 0.92 | 0.99 [0.73, 1.34] | Mot (F vs. C) | -0.04 | 0.15 | -0.26 | 0.80 | 0.96 [0.71, 1.30] |
| Mot (F vs. Ed) | -0.10 | 0.27 | -0.38 | 0.71 | 0.90 [0.53, 1.53] | Mot (F vs. Ed) | -0.09 | 0.27 | -0.32 | 0.75 | 0.92 [0.54, 1.57] |
| V.L:S.L | 0.45 | 0.29 | 1.54 | 0.12 | 1.57 [0.89, 2.77] | S.L:V.L | 0.24 | 0.27 | 0.89 | 0.37 | 1.27 [0.75, 2.17] |
| V.Q:S.L | -0.09 | 0.28 | -0.34 | 0.73 | 0.91 [0.53, 1.57] | S.Q:V.L | -0.40 | 0.27 | -1.47 | 0.14 | 0.67 [0.40, 1.14] |
| V.L:S.Q | -0.27 | 0.26 | -1.03 | 0.30 | 0.76 [0.46, 1.28] | S.L:V.Q | -0.22 | 0.25 | -0.87 | 0.39 | 0.81 [0.50, 1.31] |
| V.Q:S.Q | -0.39 | 0.32 | -1.21 | 0.22 | 0.68 [0.36, 1.27] | S.Q:V.Q | 0.57 | 0.31 | 1.81 | 0.07 | 1.77 [0.95, 3.27] |
| V.L:Age.L | **-1.26** | **0.49** | **-2.56** | **0.01*** | **0.28 [0.11, 0.74]** | S.L:Age.L | -0.07 | 0.48 | -0.15 | 0.88 | 0.93 [0.36, 2.39] |
| V.Q:Age.L | 0.30 | 0.55 | 0.55 | 0.59 | 1.35 [0.46, 3.95] | S.Q:Age.L | -0.14 | 0.46 | -0.30 | 0.76 | 0.87 [0.35, 2.14] |
| V.L:Age.Q | -0.61 | 0.32 | -1.91 | 0.06 | 0.55 [0.29, 1.02] | S.L:Age.Q | -0.33 | 0.31 | -1.05 | 0.29 | 0.72 [0.39, 1.32] |
| V.Q:Age.Q | -0.37 | 0.35 | -1.04 | 0.30 | 0.69 [0.35, 1.38] | S.Q:Age.Q | 0.26 | 0.31 | 0.85 | 0.40 | 1.30 [0.71, 2.36] |
| V.L:Day (Weekday vs. Weekend) | 0.25 | 0.22 | 1.14 | 0.25 | 1.28 [0.84, 1.97] | S.L:Day (Weekday vs. Weekend) | 0.38 | 0.20 | 1.86 | 0.06 | 1.46 [0.98, 2.17] |
| V.Q:Day (Weekday vs. Weekend) | -0.03 | 0.23 | -0.14 | 0.89 | 0.97 [0.61, 1.52] | S.Q:Day (Weekday vs. Weekend) | -0.19 | 0.22 | -0.88 | 0.38 | 0.82 [0.53, 1.27] |
| V.L:Gro (True vs. False) | 0.33 | 0.22 | 1.46 | 0.14 | 1.39 [0.89, 2.16] | S.L:Gro (True vs. False) | -0.04 | 0.24 | -0.16 | 0.87 | 0.96 [0.60, 1.53] |
| V.Q:Gro (True vs. False) | 0.14 | 0.24 | 0.58 | 0.56 | 1.15 [0.71, 1.86] | S.Q:Gro (True vs. False) | 0.36 | 0.25 | 1.47 | 0.14 | 1.44 [0.89, 2.33] |
| V.L:Bys (True vs. False) | 0.18 | 0.15 | 1.20 | 0.23 | 1.20 [0.89, 1.60] | S.L:Bys (True vs. False) | **0.31** | **0.15** | **2.12** | **0.03*** | **1.36 [1.02, 1.82]** |
| V.Q:Bys (True vs. False) | -0.11 | 0.16 | -0.70 | 0.48 | 0.89 [0.65, 1.23] | S.Q:Bys (True vs. False) | -0.06 | 0.16 | -0.36 | 0.72 | 0.95 [0.70, 1.29] |
| V.L:Ver (True vs. False) | 0.59 | 0.36 | 1.64 | 0.10 | 1.80 [0.89, 3.65] | S.L:Ver (True vs. False) | **0.70** | **0.31** | **2.30** | **0.02*** | **2.02 [1.11, 3.68]** |
| V.Q:Ver (True vs. False) | 0.27 | 0.34 | 0.81 | 0.42 | 1.31 [0.68, 2.55] | S.Q:Ver (True vs. False) | 0.51 | 0.36 | 1.40 | 0.16 | 1.66 [0.82, 3.36] |
| V.L:Mot (F vs. Ex) | -0.25 | 0.25 | -1.00 | 0.32 | 0.78 [0.47, 1.27] | S.L:Mot (F vs. Ex) | -0.05 | 0.24 | -0.21 | 0.83 | 0.95 [0.59, 1.53] |
| V.Q:Mot (F vs. Ex) | -0.41 | 0.26 | -1.57 | 0.12 | 0.66 [0.40, 1.11] | S.Q:Mot (F vs. Ex) | **-0.68** | **0.28** | **-2.46** | **0.01*** | **0.50 [0.29, 0.87]** |
| V.L:Mot (F vs. C) | -0.01 | 0.26 | -0.03 | 0.97 | 0.99 [0.60, 1.64] | S.L:Mot (F vs. C) | -0.05 | 0.25 | -0.21 | 0.83 | 0.95 [0.58, 1.56] |
| V.Q:Mot (F vs. C) | -0.00 | 0.28 | -0.01 | 0.99 | 1.00 [0.57, 1.74] | S.Q:Mot (F vs. C) | -0.44 | 0.28 | -1.56 | 0.12 | 0.64 [0.37, 1.12] |
| V.L:Mot (F vs. Ed) | 0.36 | 0.45 | 0.79 | 0.43 | 1.43 [0.59, 3.45] | S.L:Mot (F vs. Ed) | 0.28 | 0.42 | 0.68 | 0.50 | 1.33 [0.58, 3.02] |
| V.Q:Mot (F vs. Ed) | 0.59 | 0.48 | 1.22 | 0.22 | 1.80 [0.70, 4.60] | S.Q:Mot (F vs. Ed) | **1.16** | **0.52** | **2.22** | **0.03*** | **3.20 [1.15, 8.93]** |

*Model fit / thresholds (see notes):*
Q2: n=725, logLik=−610.36, AIC=1300.7; thresholds 2|3=-5.69, 3|4=-2.60, 4|5=2.30, 5|6=3.89, 6|7=4.86.
Q3: n=725, logLik=−636.96, AIC=1353.9; thresholds 2|3=-5.59, 3|4=-2.71, 4|5=1.59, 5|6=3.45, 6|7=4.53.

* indicates significant difference.